\documentstyle[12pt]{article}

  \def\pp{{\mathchoice
          {
              \kern 1pt%
              \raise 1pt
              \vbox{\hrule width5pt height0.4pt depth0pt
                    \kern -2pt
                    \hbox{\kern 2.3pt
                          \vrule width0.4pt height6pt depth0pt
                          }
                    \kern -2pt
                    \hrule width5pt height0.4pt depth0pt}%
                    \kern 1pt
           }
            {
              \kern 1pt%
              \raise 1pt
              \vbox{\hrule width4.3pt height0.4pt depth0pt
                    \kern -1.8pt
                    \hbox{\kern 1.95pt
                          \vrule width0.4pt height5.4pt depth0pt
                          }
                    \kern -1.8pt
                    \hrule width4.3pt height0.4pt depth0pt}%
                    \kern 1pt
            }
            {
              \kern 0.5pt%
              \raise 1pt
              \vbox{\hrule width4.0pt height0.3pt depth0pt
                    \kern -1.9pt  
                    \hbox{\kern 1.85pt
                          \vrule width0.3pt height5.7pt depth0pt
                          }
                    \kern -1.9pt
                    \hrule width4.0pt height0.3pt depth0pt}%
                    \kern 0.5pt
            }
            {
              \kern 0.5pt%
              \raise 1pt
              \vbox{\hrule width3.6pt height0.3pt depth0pt
                    \kern -1.5pt
                    \hbox{\kern 1.65pt
                          \vrule width0.3pt height4.5pt depth0pt
                          }
                    \kern -1.5pt
                    \hrule width3.6pt height0.3pt depth0pt}%
                    \kern 0.5pt
            }
        }}

  \def\mm{{\mathchoice
                       {
                             \kern 1pt
               \raise 1pt    \vbox{\hrule width5pt height0.4pt depth0pt
                                  \kern 2pt
                                  \hrule width5pt height0.4pt depth0pt}
                             \kern 1pt}
                       {
                            \kern 1pt
               \raise 1pt \vbox{\hrule width4.3pt height0.4pt depth0pt
                                  \kern 1.8pt
                                  \hrule width4.3pt height0.4pt depth0pt}
                             \kern 1pt}
                       {
                            \kern 0.5pt
               \raise 1pt
                            \vbox{\hrule width4.0pt height0.3pt depth0pt
                                  \kern 1.9pt
                                  \hrule width4.0pt height0.3pt depth0pt}
                            \kern 1pt}
                       {
                           \kern 0.5pt
             \raise 1pt  \vbox{\hrule width3.6pt height0.3pt depth0pt
                                  \kern 1.5pt
                                  \hrule width3.6pt height0.3pt depth0pt}
                           \kern 0.5pt}
                       }}

\catcode`@=11
\def\un#1{\relax\ifmmode\@@underline#1\else
        $\@@underline{\hbox{#1}}$\relax\fi}
\catcode`@=12

\let\du=\du                     

\def\a{\alpha}
\def\b{\beta}
\def\c{\chi}
\def\d{\delta}
\def\e{\epsilon}
\def\f{\phi}
\def\g{\gamma}
\def\h{\eta}

\def\j{\psi}
\def\k{\kappa}
\def\l{\lambda}
\def\m{\mu}
\def\n{\nu}
\def\o{\omega}
\def\p{\pi}
\def\q{\theta}
\def\r{\rho}
\def\s{\sigma}
\def\t{\tau}

\def\z{\zeta}
\def\D{\Delta}

\def\G{\Gamma}

\def\L{\Lambda}
\def\O{\Omega}

\def\S{\Sigma}

\def\ve{\varepsilon}
\def\vf{\varphi}

\def\vq{\vartheta}

\def\cc{{\cal C}}

\def\ck{{\cal K}}

\def\cm{{\cal M}}

\def\cy{{\cal Y}}
\def\cz{{\cal Z}}

\def\bo{{\raise-.5ex\hbox{\large$\Box$}}}               
\def\pa{\partial}                                       
\def\de{\nabla}                                         
\def\pr{\prod}                                          
\def\TH{{\raise.2ex\hbox{$\displaystyle \bigodot$}\mskip-4.7mu \llap H \;}}
\def\face{{\raise.2ex\hbox{$\displaystyle \bigodot$}\mskip-2.2mu \llap {$\ddot
        \smile$}}}                                      

\def\sp#1{{}^{#1}}                              
\def\VEV#1{\left\langle #1\right\rangle}        
\def\abs#1{\left| #1\right|}                    
\def\leftrightarrowfill{$\mathsurround=0pt \mathord\leftarrow \mkern-6mu
        \cleaders\hbox{$\mkern-2mu \mathord- \mkern-2mu$}\hfill
        \mkern-6mu \mathord\rightarrow$}
\def\dvec#1{\vbox{\ialign{##\crcr
        \leftrightarrowfill\crcr\noalign{\kern-1pt\nointerlineskip}
        $\hfil\displaystyle{#1}\hfil$\crcr}}}           
\def\dt#1{{\buildrel {\hbox{\LARGE .}} \over {#1}}}     

\def\frac#1#2{{\textstyle{#1\over\vphantom2\smash{\raise.20ex
        \hbox{$\scriptstyle{#2}$}}}}}                   
\def\sfrac#1#2{{\vphantom1\smash{\lower.5ex\hbox{\small$#1$}}\over
        \vphantom1\smash{\raise.4ex\hbox{\small$#2$}}}} 
\def\bfrac#1#2{{\vphantom1\smash{\lower.5ex\hbox{$#1$}}\over
        \vphantom1\smash{\raise.3ex\hbox{$#2$}}}}       
\def\afrac#1#2{{\vphantom1\smash{\lower.5ex\hbox{$#1$}}\over#2}}    

\def\[{\lfloor{\hskip 0.35pt}\!\!\!\lceil}
\def\]{\rfloor{\hskip 0.35pt}\!\!\!\rceil}
\def\Lag{{\cal L}}
\def\du#1#2{_{#1}{}^{#2}}

\def\fracm#1#2{\hbox{\large{${\frac{{#1}}{{#2}}}$}}}

\def\un{\underline}
\def\fracmm#1#2{{{#1}\over{#2}}}

\def\low#1{{\raise -3pt\hbox{${\hskip 0.75pt}\!_{#1}$}}}

\def\Dot#1{\buildrel{_{_{\hskip 0.01in}\bullet}}\over{#1}}
\def\dt#1{\Dot{#1}}

\newskip\humongous \humongous=0pt plus 1000pt minus 1000pt
\def\caja{\mathsurround=0pt}
\def\eqalign#1{\,\vcenter{\openup2\jot \caja
        \ialign{\strut \hfil$\displaystyle{##}$&$
        \displaystyle{{}##}$\hfil\crcr#1\crcr}}\,}
\newif\ifdtup

\def\ref#1{$\sp{#1)}$}

\def\pl#1#2#3{Phys.~Lett.~{\bf {#1}B} (19{#2}) #3}
\def\np#1#2#3{Nucl.~Phys.~{\bf B{#1}} (19{#2}) #3}
\def\prl#1#2#3{Phys.~Rev.~Lett.~{\bf #1} (19{#2}) #3}
\def\pr#1#2#3{Phys.~Rev.~{\bf D{#1}} (19{#2}) #3}
\def\cqg#1#2#3{Class.~and Quantum Grav.~{\bf {#1}} (19{#2}) #3}
\def\cmp#1#2#3{Commun.~Math.~Phys.~{\bf {#1}} (19{#2}) #3}

\def\ijmp#1#2#3{Int.~J.~Mod.~Phys.~{\bf A{#1}} (19{#2}) #3}

\parskip=\medskipamount                 
\thicklines                         

\begin{document}
\thispagestyle{empty}

{\hbox to\hsize{
\vbox{\noindent KL~--~TH 01/01   \hfill February 2001 \\
hep-th/0102099 }}}

\noindent
\vskip1.3cm
\begin{center}

{\Large\bf Universal Hypermultiplet Metrics}\\

\vglue.3in

Sergei V. Ketov 
\footnote{Supported in part by the `Deutsche Forschungsgemeinschaft'}

{\it Department of Theoretical Physics\\
     Erwin Schr\"odinger Strasse \\
     University of Kaiserslautern}\\
{\it 67653 Kaiserslautern, Germany}
\vglue.1in
{\sl ketov@physik.uni-kl.de}
\end{center}

\vglue.3in

\begin{center}
{\Large\bf Abstract}
\end{center}

\noindent
Some instanton corrections to the universal hypermultiplet moduli space metric 
of the type-IIA string theory compactified on a Calabi-Yau threefold arise due 
to multiple wrapping of BPS membranes and fivebranes around certain cycles of 
Calabi-Yau. The classical universal hypermultipet metric is locally equivalent
 to the Bergmann metric of the symmetric quaternionic space  $SU(2,1)/U(2)$, 
whereas its generic quaternionic deformations are governed by the integrable 
$SU(\infty)$ Toda equation. We calculate the exact (non-perturbative) UH 
metrics in the special cases of (i) the D-instantons (the wrapped D2-branes) 
in the absence of fivebranes, and (ii) the fivebrane instantons with vanishing
 charges, in the absence of D-instantons. The solutions of the first type 
preserve the $U(1)\times U(1)$ classical symmetry, while they can be 
interpreted as the gravitational dressing of the hyper-K\"ahler D-instanton 
solutions. The solutions of the second type preserve the non-abelian $SU(2)$ 
classical symmetry, while they can be interpreted as the gradient flows in the
 universal hypermultiplet moduli space. 

\newpage

\section{Introduction}

Non-perturbative (instanton) quantum corrections in compactified 
M-theory/strings are believed to be crucial for solving the fundamental
problems of vacuum degeneracy and supersymmetry breaking \cite{strings}. Some 
instanton corrections to various physical quantities in the effective 
lower-dimensional supergravity theory, originating from the string/M-theory 
compactification  on a {\it Calabi-Yau} (CY) threefold $\cy$, can be 
understood in terms of the Euclidean five-branes wrapped about the entire CY 
space and the Euclidean membranes (two-branes) wrapped about special 
(supersymmetric) three-cycles $\cc$ of $\cy$ \cite{bbs,bb}.~\footnote{The 
supersymmetric 3-cycle $\cc$ is a three-dimensional submanifold of the
CY space, which obeys 
$\left.J\right|_{\cc}=\left.{\rm Im}(\O)\right|_{\cc}=0$, where $J$ is
 the K\"ahler form and $\O$ is the holomorphic $(3,0)$ form in $\cy$. The
supersymmetric cycles minimize volume in their homology class \cite{ka}, 
while the corresponding wrapped brane configurations lead to the BPS states.} 
Being solitonic (BPS) classical solutions to the higher dimensional 
(Euclidean) equations of motion with non-vanishing topological charges, these 
wrapped branes are localized in the uncompactified dimensions and thus can be 
identified with the instantons. The instanton actions are essentially given by 
the volumes of the cycles on which the branes are wrapped.
 
For instance, the compactification of the type-IIA superstring theory on  
$\cy$ gives rise to the four-dimensional (4d) N=2 superstrings whose 
{\it Low-Energy Effective Action} (LEEA) is given by the 4d, N=2
supergravity coupled to N=2 vector supermultiplets and hypermultiplets. 
Any other N=2 matter multiplet (of physical spin $\leq 1$ or $\leq 1/2$) can 
be dualized to an N=2 vector multiplet or a hypermultiplet, respectively. 
The numbers of the N=2 matter vector multiplets and hypermultiplets are 
dictated by the topological data about $\cy$, namely, the Hodge numbers 
$h_{1,1}$ and $h_{1,2}+1$, respectively, where $h_{p,q}$ are the dimensions 
of the Dolbeaux cohomology groups of $\cy$ \cite{strings}. The hypermultiplet 
LEEA is most naturally described by the {\it Non-Linear Sigma-Model} (NLSM), 
whose scalar fields parametrize the quaternionic target space $\cm_H$ 
\cite{bw}. The instanton corrections to the LEEA due to the wrapped 
fivebranes and membranes can be easily identified and distinguished from each 
other in the semi-classical limit, since the fivebrane instanton corrections 
are organized by powers of $e^{-1/g^2_{\rm string}}$, whereas the membrane 
instanton corrections are given by powers of $e^{-1/g_{\rm string}}$, where 
$g_{\rm string}$ is the type-IIA superstring coupling constant \cite{w1}.

{}From the M-theory perspective, the compactified field theory (LEEA) is given
by the five-dimensional supergravity with the same number of unbroken 
supercharges. The effective supergravities in four and five dimensions are 
related via the compacification of the latter on a circle, which does not 
affect the NLSM target space $\cm_H$ in the hypermultiplet sector. The
vacuum expectation value of the four-dimensional dilaton field $\VEV{\f}$ in
 the compacitified type-IIA superstring is simply related to the CY volume 
$V_{\rm CY}$ in M-theory, $V_{\rm CY}=e^{-2\VEV{\f}}$, so that the type-IIA 
superstring loop expansion amounts to the derivative expansion of the M-theory
action \cite{one}.

The unbroken local supersymmetry with eight supercharges (e.g., N=2 in 4d) put 
severe constraints on the LEEA of matter supermultiplets or, equivalently, on 
their moduli space. In fact, it requires the whole moduli space $\cm$ be the 
local product, $\cm=\cm_V\times \cm_H$, where $\cm_V$ is the special
 K\"ahler manifold associated with the N=2 vector multiplets \cite{wp},
 and $\cm_H$ is the quaternionic manifold associated with the hypermultiplets
\cite{bw}.~\footnote{See e.g., ref.~\cite{aspin} for a review.} Any CY 
compactification has the co-called {\it Universal Hypermultiplet} (UH) 
containing a dilaton, an axion, a complex RR-type pseudo-scalar and a Dirac 
dilatino \cite{fcg}. The constraints imposed by 4d, N=2 local supersymmetry on 
the four-dimensional UH moduli space are even stronger then in higher 
dimensions: the target space of the universal hypermultiplet NLSM has to be an
 Einstein space with the (anti)self-dual Weyl tensor (these spaces are called 
the self-dual Einstein spaces or simply the Einstein-Weyl spaces in the 
mathematical literature).

The N=2 vector multiplet moduli space $\cm_V$ is relatively well understood, 
whereas much less is known about the hypermultiplet moduli space 
$\cm_H$ \cite{aspin}. The conjectured duality between the type-IIA superstring
 compactification on CY and the heterotic string compactification on 
$K3\times T^2$, which is supposed to exchange $\cm_V$ and $\cm_H$ in 
{\it the same} moduli space $\cm$, was used in the past to get information 
about $\cm_H$ on the type-II string side from the knowledge about $\cm_V$ on 
the heterotic string side, although with only partial success \cite{h2d}. 
Since the UH transforms into the N=2 supergravity multiplet under type-II 
mirror symmetry, any quantum corrections to the classical UH metric are 
essentially gravitational in nature \cite{one}.

The standard instanton calculus gives us another technical device after 
taking into account the partial supersymmetry breaking induced by the BPS 
branes \cite{bbs}. Being applied to the bosonic instanton background, 
broken supersymmetries generate fermionic (Goldstino) zero modes that are to 
be absorbed by extra terms in the effective field theory. These 
instanton-induced interactions are quartic in the fermionic fields (by index
theorems) and thus contribute to the curvature tensor of the supersymmetric 
NLSM. By supersymmetry, the instanton corrections to the NLSM curvature imply 
the corresponding deformations in the NLSM metric. To calculate these 
non-perturbative corrections by using the standard instanton technology, one 
may integrate over the fermionic zero modes and then compute the fluctuation 
determinants, which is apparently the hard problem \cite{hm}. In this paper we
 restrict ourselves to a calculation of the instanton corrections to the 
universal hypermultiplet NLSM metric, by analyzing generic quaternionic
deformations of the classical UH metric. We apply the most direct procedure 
based on the fact that the (anti)self-dual Weyl tensor already implies the 
integrable system of partial differential equations on the components of the 
UH moduli space metric. Additional simplifications arise due to the Einstein
condition and the physically motivated isometries. The UH metric in question 
is supposed to be regular and complete.

The local data about the underlying CY threefold $\cy$ is described in terms 
of the $h_{p,q}$ complex moduli. In the compactified theory, the $h_{1,2}$ 
moduli are promoted to the non-universal matter hypermultiplets. The universal
 hypermultiplet is `universal' in the sense that its own moduli space does not
 depend upon the CY moduli, or, equivalently, the UH is merely sensitive to the
global information about the CY space (like topology, volume, charges and 
symmetries).

Some instanton solutions to the effective Euclidean (UH + supergravity) 
equations of motion and the corresponding instanton actions were found in  
refs.~\cite{bb,gs,gs2}. The instanton solutions carry charges descending from 
the wrapped brane charges, and they preserve a part (half) of type-IIA 
supersymmetry. The instanton actions are also dependent upon the charges of 
the instanton and the complex structure (or central charge) at the moduli 
space infinity. Unfortunately, no exact quaternionic metrics in the UH moduli 
space beyond string perturbation theory were constructed, which would amount to
a calculation of the exact LEEA of the universal hypermultiplet. In this paper
we report about the results of our investigation initiated in ref.~\cite{din}.

The paper is organized as follows. In sect.~2 we review the four-dimensional 
classical LEEA (or NLSM) of the universal hypermultiplet in the background of 
4d, N=2 supergravity \cite{fsh}, and discuss its perturbative deformations 
\cite{one,ger} and the origin of non-perturbative corrections due to the 
wrapped BPS branes \cite{bbs,bb,gs,gs2}. The symmetry structure of the 
classical UH moduli space is given in sect.~3. The D-instanton corrections due
 to the wrapped membranes (D2-branes) in the hyper-K\"ahler limit \cite{ova} 
are presented in sect.~4, whereas their quaternionic generalizations are 
described in sect.~5. In sect.~6 we demonstrate that the zero-charge fivebrane
 instanton corrections to the UH metric are described by the Tod-Hitchin 
metric. Its physical interpretation as the gradient flow in the UH moduli 
space is also given in sect.~6. Generic UH metrics are briefly discussed in 
sect.~7. In Appendix A we collect basic facts about the quaternionic symmetric
 space $SU(2,1)/U(2)$. Appendix B is devoted to basic definitions of theta 
functions and some useful identities between them.

\section{M-theory, type-IIA string perturbation theory, 
         and the UH classical moduli space metric}

The eleven-dimensional (11d) M-theory is supposed to substitute the 
ten-dimensional (10d) type-IIA superstring theory at strong string coupling, 
while the M-theory LEEA is described by 11d supergravity \cite{wit}. 
The action of 11d supergravity is unique, and its bosonic part 
reads \cite{ecole}~\footnote{The 11d signature is $(-,+,+,\ldots,+)$.}
$$ S_{11}= \fracmm{1}{2\k^2_{11}}\int d^{11}x\sqrt{-\hat{g}}\left(
\hat{\cal R} -\fracm{1}{48}\hat{F}^2\right) + \fracmm{1}{12\k^2_{11}}\int
\hat{A}_3\wedge \hat{F}_4\wedge \hat{F}_4~~,\eqno(2.1)$$
where  $\k_{11}$ is the 11d gravitational constant, $\hat{F}_4=d\hat{A}_3$,
and all quantities with `hats' refer to 11d. Though the 11d supergravity 
action is only the leading (LEEA) part of the full (unknown) M-theory action, 
it is already enough to study the BPS states in M-theory. The latter are given
 by the solitonic classical solutions (called M-branes) to 11d supergravity, 
which preserve some part (say, a half) of 11d supersymmetry. The BPS condition
 amounts to the existence of a 
Majorana-Killing 11d spinor $\ve_{\a}(x)$ obeying the first-order linear
differential equation
$$ \tilde{D}_M\ve\equiv\left[ \left( \pa_M+\fracm{1}{4}\o\du{M}{BC}\G_{BC}
\right) - \fracm{1}{288}\left( \G\du{M}{NPQS}-8\d^N_M \G^{PQS}\right)
F_{NPQS}\right]\ve=0~,\eqno(2.2)$$
where $M=0,1,\ldots,10$, $\o\du{M}{BC}$ is the 11d supergravity spin 
connection, and $\G$'s are the antisymmetric products (with unit weight) of 
11d Dirac matrices. The M-theory supersymmetry algebra is just the most 
general super-Poincar\'e algebra in 11d \cite{town},
$$ \{Q_{\a},Q_{\b}\}=(C\G^M)_{\a\b}P_M +\fracm{1}{2!}(C\G^{MN})_{\a\b}Z_{MN}
+ \fracm{1}{5!}(C\G^{MNPQS})_{\a\b}Y_{MNPQS}~,\eqno(2.3)$$ 
where $Q_{\a}$, $\a=1,\ldots,32$, are the 11d supersymmetry charges,
$P_M$ are 11d translations, $C$ is the 11d charge conjugation matrix, whereas
$Z_{MN}$ and $Y_{MNPQS}$ stand for the `electric' and `magnetic' charges,
respectively,
$$ \eqalign{
Z_{MN}~=~&\int_{\rm M2-brane} dx^M\wedge dx^N~,\cr
Y_{MNPQS}~=~&\int_{\rm M5-brane} dx^M\wedge dx^N \wedge dx^P \wedge dx^Q
\wedge dx^S~.\cr}\eqno(2.4)$$
Equation (2.3) is just the 11d Fierz rearrangement formula for 
$\{Q_{\a},Q_{\b}\}\in\underline{528}$ of $Spin(32)$. The M-theory
solitons are thus given by membranes \cite{bst} and fivebranes \cite{guv}, 
which are in fact dual to each other  under the `electric-magnetic' duality 
in 11d.

The physical and topological significance of the abelian 
(`electric' and `magnetic') charges in eq.~(2.3) become clear from the 11d
equations of motion, which follow from the action (2.1). In particular, one 
has 
$$ d\,{}^*_{11}\!\hat{F}_4+\fracm{1}{2}\hat{F}_4^2=d\left( {}^*_{11}\hat{F}_4
  + \fracm{1}{2}\hat{A}_3\wedge\hat{F}_4\right)=0~,\eqno(2.5)$$
where ${}^*_{11}\hat{F}_4$ is the 11d dual to $\hat{F}_4$. Hence, the 
`electric' charge of an M2-brane \cite{pag},
$$ Q_{\rm e}=\fracmm{1}{\k_{11}\sqrt{2}}\int_{S^7}\left(
 {}^*_{11}\hat{F}_4+ \fracm{1}{2}\hat{A}_3\wedge\hat{F}_4\right)~,\eqno(2.6)$$
is conserved, where $S^7$ is the asymptotic seven-sphere surrounding the 
M2-brane ({\it cf.} Gauss law in Maxwell electrodynamics). 
Similarly, the `magnetic' charge of the M5-brane \cite{pag},
$$ Q_{\rm m}= \fracmm{1}{\k_{11}\sqrt{2}}\int_{S^4} \hat{F}_4~,\quad
d\hat{F}_4=0~,\eqno(2.7)$$
is also conserved, where $S^4$ is the asymptotic four-sphere surrounding the 
M5-brane ({\it cf.} Dirac monopole charge). The Dirac quantization condition 
implies 
$$  Q_{\rm e} Q_{\rm m}=2\p(\hbar c){\bf Z}~.\eqno(2.8)$$

Demanding the membrane and fivebrane (worldvolume) actions be well defined
in quantum theory gives rise to a quantization of their electric and magnetic
charges or, equivalently, the M-brane tension quantization \cite{tens,tens2},
$$  2\p T_5= T^2_2 \quad {\rm and}\quad 
\k_{11}^2T_2T_5=\p{\bf Z}~.\eqno(2.8)$$

A connection to type-IIA superstrings is obtained by compactifying 
M-theory/11d supergravity on a cirle of radius $r=\sqrt{\a'}$ with the
{\it Kaluza-Klein} (KK)  Ansatz \cite{wit}
$$ d\hat{s}^2_{11}=e^{-2\f/3}d\tilde{s}^2_{10}+e^{4\f/3}(dx^{11}-A_mdx^m)^2~,
\eqno(2.9)$$  
where $d\tilde{s}^2_{10}=\tilde{g}_{mn}dx^mdx^n$ is the 10d spacetime metric 
(in the string frame) with $m,n=0,1,\ldots,9$, $A_m$ is the KK (type-IIA) 
vector field, and $\f$ is the 10d dilaton. The gravitational constants in 
11d and 10d are related by $\k_{10}^2=\k_{11}^2/(2\p\sqrt{\a'})$. Performing 
the dimensional reduction of the action
(2.1) on the circle with the help of eq.~(2.9) gives rise to the bosonic part
of the type-IIA supergravity action in the string frame. The latter is related
 to the canonical form of the type-IIA supergravity action (in the Einstein 
frame) via the Weyl rescaling $\tilde{g}_{mn}= e^{\f/2}g_{mn}$. This yields 
the standard bosonic part of the 10d type-IIA supergravity action,
$$\eqalign{
S_{10}~=~& \fracmm{1}{2\k^2_{10}}\int d^{10}x\sqrt{-g}\left[
{\cal R} -\fracm{1}{2}(\pa_m\f)^2\right]-
 \fracmm{1}{4\k^2_{10}}\int \left[ e^{\f/2}F_4\wedge{}^*_{10}F_4 +
e^{-\f}H_3\wedge {}^*_{10}H_3 \right]\cr
~& +\fracmm{1}{4\k^2_{10}}\int B_2\wedge F_4\wedge F_4~~,\cr}\eqno(2.10)$$
where $H_3=dB_2$ and $F_4=dA_3$ originate from the three-form $\hat{A}_3$ and 
its field strength $\hat{F}_4$ in 11d, 
e.g., $\hat{A}_3=A_3+B_2\wedge dx^{11}$, etc.

The 10d type-IIA supergravity is known to be the LEEA of 10d type-IIA 
strings in the (string) tree approximation \cite{strings}. The double 
dimensional reduction of the 11d (M-theory) membrane yields a (type-IIA) 
fundamental string in 10d. The fundamental string in 10d is dual to a solitonic
(NS-NS) five-brane. In M-theory compactified on the circle $S^1$ the 11d
fivebrane can be wrapped about $S^1$, which gives rise to the (RR-charged)
type-IIA fourbrane called D4-brane \cite{town}. In 10d this D4-brane is dual 
to the 10d (RR-charged) membrane called D2-brane \cite{town}. 

The non-perturbative (instanton) corrections to the four-dimensional UH 
originate from the 10d type-IIA membranes (D2-branes) wrapped about certain 
(supersymmetric or special Lagrangian) 3-cycles $\cc$ of CY, and the 10d 
type-IIA NS-NS  fivebranes wrapped about the entire CY space 
\cite{bbs,bb,gs,gs2}. From the 11d M-theory perspective, the dilaton 
expectation value is related to the volume of the CY space (sect.~1), whereas 
the expectation value of the RR scalar $C$ is related to the CY period 
$\int_{\cc}\O$.

The universal sector (UH) of the CY compactification of the 10d type-IIA 
supergravity in four dimensions is most easily obtained by using the following 
{\it Ansatz} for the 10d metric \cite{bb}: 
$$ ds^2_{10}=g_{mn}dx^mdx^n
=e^{-\f/2}ds^2_{\rm CY}+e^{3\f/2}g_{\m\n}dx^{\m}dx^{\n}~,\eqno(2.11)$$
while keeping only $SU(3)$ singlets in the internal CY indices \cite{wit2}
and ignoring all CY complex moduli. In eq.~(2.11) $\f(x)$ stands for the 4d 
dilaton, $g_{\m\n}(x)$ is the spacetime
metric in four uncompactified dimensions, $\m,\n=0,1,2,3$, and 
$ds^2_{\rm CY}$ is the (K\"ahler and Ricci-flat) metric of the internal CY 
threefold $\cy$ in complex coordinates, 
$$ds^2_{\rm CY}=g_{i\bar{j}}(y,\bar{y})dy^id\bar{y}^{\bar{j}}~,\eqno(2.12)$$
where $i,j=1,2,3$. By definition, the CY threefold $\cy$ possesses the $(1,1)$ 
K\"ahler form $J$ and the holomorphic $(3,0)$ form $\O$. The universal 
hypermultipet (UH) unites the dilaton $\f$, the axion $D$ coming from 
dualizing the three-form field strength $H_3=dB_2$ of the NS-NS two-form $B_2$ 
in 4d, and the complex scalar $C$ representing the RR three-form $A_3$ with 
$A_{ijk}(x,y) =\sqrt{2}C(x)\O_{ijk}(y)$. When using a flat (or rigid) CY 
({\it cf.} a six-torus of equal radii $r=\sqrt{\a'}$), with 
$$  g_{i\bar{j}}=\d_{i\bar{j}}\quad {\rm and}\quad \O_{ijk}=\ve_{ijk}~,
\eqno(2.13)$$
the reduction of eq.~(2.10) down to four dimensions with the help of 
eqs.~(2.12) and (2.13) yields the so-called Ferrara-Sabharwal action 
\cite{fsh}:~\footnote{The 4d spacetime signature is $(-,+,+,+)$. The 4d 
gravitational constant $\k_4$ is related to the \newline ${~~~~~}$ 10d 
gravitational constant $\k_{10}$ via the equation
 $\k^2_4=\k^2_{10}/(2\p r)^6=\k^2_{10}/(2\p\sqrt{\a'})^6$.}
$$\eqalign{
S_{4}~=~& \fracmm{1}{2\k^2_{4}}\int d^{4}x\sqrt{-g}\left[
{\cal R} -2(\pa_m\f)^2 -2e^{2\f}\abs{\pa_{\m}C}^2\right] \cr
~& - \fracmm{1}{4\k^2_{4}}\int \left[ e^{-4\f}H_3\wedge{}^*_{4}H_3 -
2iH_3\wedge \bar{C}\dvec{d}C-4H_3\wedge dD\right]~,\cr}\eqno(2.14)$$
where the real 4d Lagrange multiplier $D$ has been introduced to enforce the 
Bianchi relation $dH_3=0$. Removing $H_3$ via its
equations of motion from eq.~(2.14) results in the dual action \cite{bb}
$$
S_{4}=-\fracmm{1}{\k^2_{4}}\int d^{4}x\sqrt{-g}\left[ -\fracm{1}{2}{\cal R}
 + (\pa_m\f)^2 +e^{2\f}\abs{\pa_{\m}C}^2 +e^{4\f}\left(\pa_{\m}D+
\fracm{i}{2}\bar{C}\dvec{\pa_{\m}}C\right)^2\right]~.\eqno(2.15)$$

The equivalent action in the string frame reads \cite{ger} 
$$ \eqalign{
S_{\rm 4,~string} ~=~&  \fracmm{1}{\k^2_{4}}\int d^4x\sqrt{-\tilde{g}}\,
e^{-2\f}\left\{ \fracm{1}{2}\tilde{\cal R} 
+2(\pa_{\m}\f)^2-\fracm{1}{6}H^2_{\m\n\l}\right\} \cr
~& - \fracmm{1}{\k^2_{4}}\int d^4x\sqrt{-\tilde{g}}\left[ 
\pa_{\m}C\pa^{\m}\bar{C}+
\fracm{i}{2}H^{\m}\left(C\pa_{\m}\bar{C}-\bar{C}\pa_{\m}C\right)\right]~,\cr}
\eqno(2.16)$$
where $H^{\r}=\fracmm{1}{6\sqrt{-g}}\ve^{\r\m\n\l}H_{\m\n\l}$ is the Hodge 
dual of the field strength $H_3=dB_2$. The N=2 supergravity background in 
eqs.~(2.14), (2.15) and (2.16)  is merely represented by the 4d spacetime 
metric. The NLSM action of the UH thus has the following target space
metric \cite{fsh}:
$$ -\k^2_4 ds^2_{\rm classical}=d\f^2+e^{2\f}\abs{dC}^2+e^{4\f}\left(dD
+\fracm{i}{2}\bar{C}\dvec{d}C\right)^2~, \eqno(2.17)$$
whose right-hand-side is regular and positively definite. This metric is of 
purely gravitational origin, with $\k^2_4$ being the only coupling constant.

The perturbative (one-loop) string corrections to the UH metric (2.17) 
originate from the $(Riemann)^4$ terms \cite{r4} in M-theory compactified on a
 CY three-fold $\cy$ \cite{one}. These quantum corrections are proportional to
 the CY Euler number $\c=2\left(h_{1,1}-h_{1,2}\right)$ \cite{cand}. 
The one-loop corrected action of UH coupled to gravity in the 
{\it string frame} reads \cite{ger} 
$$ \eqalign{
S_{\rm 4,~one-loop} ~=~& \fracmm{1}{\k^2_{4}}
\int d^4x \sqrt{-\tilde{g}}\left\{ 
\left(e^{-2\f}+\hat{\c}\right) \left( \fracm{1}{2}{\cal R} 
 -\fracm{1}{6}H^2_{\m\n\l}\right) 
+\fracmm{2e^{-4\f}}{e^{-2\f}+\hat{\c}}(\pa_{\m}\f)^2\right\}\cr
~& - \fracmm{1}{\k^2_{4}}\int d^4x\sqrt{-\tilde{g}}\left[ 
\pa_{\m}C\pa^{\m}\bar{C}+
\fracm{i}{2}H^{\m}\left(C\pa_{\m}\bar{C}-\bar{C}\pa_{\m}C\right)\right]~,\cr}
\eqno(2.18)$$
where $\hat{\c}$ is proportional to $\c$.~\footnote{The
relative coefficient varies in the literature, see e.g., ref.~\cite{one}.}  
The superstring (loop) perturbation theory is controlled by the powers of 
$e^{-2\f}$, whereas the second line of eq.~(2.18) cannot be multiplied by any 
power of $e^{-2\f}$ without breaking the  perturbative Peccei-Quinn type 
symmetry $C\to C+const$. This non-renormalization argument does not, however, 
exclude the non-perturbative (gravitational-type) corrections due to the 
wrapped branes, which break the Peccei-Quinn type symmetries (sects.~4 and 5). 

The string one-loop corrected NLSM metric dictated by eq.~(2.18) is related to
 the classical metric of eq.~(2.17) by a local field redefinition,
$e^{-2\f}\to e^{-2\f}+\hat{\c}$, so that the local UH geometry is not affected 
by perturbtaive string corrections \cite{one}. String duality, however, implies
some discrete (global) identifications in the UH moduli space, which are the
consequence of the brane charge (and tension) quantization in M-theory 
\cite{bb} (see the next sect.~3).  

\section{Classical metric and perturbative deformations} 

The NLSM of UH with the target space metric (2.17) can be rewritten to the form
 \cite{bbs,bb} 
$$ \Lag_{\rm NLSM}= -\,\fracmm{1}{\k^2_4}\left(
\ck_{,S\bar{S}}\pa_{\m}S\pa^{\m}\bar{S}+
\ck_{,S\bar{C}}\pa_{\m}S\pa^{\m}\bar{C}+
\ck_{,C\bar{S}}\pa_{\m}C\pa^{\m}\bar{S}+
\ck_{,C\bar{C}}\pa_{\m}C\pa^{\m}\bar{C}\right) \eqno(3.1)$$
in terms of two complex variables, $C$ and $S$,
$$ S = e^{-2\f}+2iD+C\bar{C}~, \eqno(3.2)$$
where commas in subscripts denote partial derivatives, and 
the K\"ahler potential reads
$$ \ck= -\log\left( S +\bar{S}-2C\bar{C}\right)= -\log\left(2e^{-2\f}\right)~.
\eqno(3.3)$$
Equation (3.1) makes manifest the K\"ahler nature of the classical NLSM 
metric,  
$$ ds^2_{\rm NLSM}= e^{2\ck}\left[dSd\bar{S}-2CdSd\bar{C}-2\bar{C}d\bar{S}dC
+2(S+\bar{S})dCd\bar{C}\right]~.\eqno(3.4)$$
Other useful parametrizations are given in Appendix A. In particular, after
the change of variables (A.9) or its inverse,
$$  S = \fracmm{1-z_1}{1+z_1}~,\qquad C=\fracmm{z_2}{1+z_1}~,\eqno(3.5)$$
and the K\"ahler gauge transformation (A.4) with 
$f(z)=\log[\fracm{1}{2}(1+z_1)]$, the K\"ahler potential (3.3) takes the form 
(A.3) that is associated with the standard (Bergmann) metric (A.1) of the 
non-compact symmetric space $SU(2,1)/SU(2)\times U(1)$. When using the new
coordinates (A.5), one arrives at the classical UH metric in the form (A.6), 
with the $SU(2)$ isometry of the metric being manifest. 

The classical quaternionic space $Q=G/H\equiv SU(2,1)/SU(2)\times U(1)$ 
has the eight-dimensional non-abelian isometry group $G=SU(2,1)$ whose left
action on $Q$ after the compensating right action of the `gauge' subgroup 
$H=SU(2)\times U(1)$ leaves the metric (A.1) of $Q$ intact. 

As far as quantum corrections within the type-IIA string (loop) perturbation 
theory are concerned, the UH metric is supposed to be invariant under the 
so-called Peccei-Quinn type symmetries \cite{bbs,one}. These symmetries are 
given by constant shifts of the NS-NS axion $D$,
$$ S\to S + i\h~,\eqno(3.6)$$
and constant shifts of the R-R field $C$~\cite{dew},
$$ C\to C+ \g-i\b~,\quad  S\to S + 2(\g+i\b)C+\g^2+\b^2~,\eqno(3.7)$$
where $\h$, $\b$ and $\g$ are the real parameters. The Peccei-Quinn type 
transformations (3.6) and (3.7) form the non-abelian Heisenberg group 
\cite{bb}. It is usually assumed \cite{bbs,bb,one} that the Peccei-Quinn type
symmetries do not change their form in string perturbation theory. Since our
considerations are purely local, we ignore the global group structure and
refer to the symmetry (3.6) as $U_D(1)$.

The instantons originating from the fivebranes wrapped over the entire CY 
space break both symmetries (3.6) and (3.7) ({\it cf.} the breaking of the 
translational invariance of the $\theta$-parameter in QCD by instantons), 
whereas the D-instantons (wrapped membranes) break the symmmetry (3.7) but
keep the symmetry (3.6) \cite{ova}.

The $U(1)$ subgroup of the $SU(2)\subset H$ symmetry is given by the duality 
rotations $U_C(1)$ of the complex R-R pseudo-scalar $C$,
$$ C\to e^{ i\a} C~,\qquad S\to S~, \eqno(3.8)$$
with the real parameter $\a$. The duality rotations (3.8) are believed to be 
exact in quantum theory, even when all quantum (or instanton) corrections  
are taken into account \cite{bbs,bb}. 

The remaining four clasical symmetries of the coset $SU(2,1)/U(2)$  are given 
by scale transformations,
$$ S\to S+ \l S~,\quad  C\to C +\fracm{1}{2}\l C~,\eqno(3.9)$$
with the real parameter $\l$, and \cite{dew}
$$\eqalign{
 S~\to~& S+ \fracm{i}{4}\e_1 S^2+\fracm{1}{2}(\e_2+i\e_3)CS~,\cr
 C~\to~& C+ \fracm{i}{4}\e_1 CS+\fracm{1}{2}\e_2
(C^2-\fracm{1}{2}S) +\fracm{i}{2}\e_3(C^2+\fracm{1}{2}S)~,\cr}\eqno(3.10)$$
with the real parameters $\e_1$, $\e_2$ and $\e_3$. The symmetries (3.9) and
(3.10) are always broken by instanton corrections.

The conserved Noether charges associated with the isometries (3.6) and (3.7) 
were calculated in ref.~\cite{bb},
$$ \eqalign{
J_{\h}~=~&\fracmm{i}{\k^2_{4}}e^{2\ck}\left(dS-d\bar{S}+2C\dvec{d}\bar{C}
\right)~,\cr
J_{\b}~=~&-\,\fracmm{2i}{\k^2_{4}}e^{\ck}\left(dC-d\bar{C}\right)+2\left(
C+\bar{C}\right)J_{\h}~,\cr 
J_{\g}~=~&-\,\fracmm{2}{\k^2_{4}}e^{\ck}\left(dC+d\bar{C}\right)-2i\left(
C-\bar{C}\right)J_{\h}~.\cr}\eqno(3.11)$$
The Noether charge (over a supersymmetric three-cycle $\cc$)
$$ Q_{\h}= \int_{\cc} {}^*_{4}J_{\h}\,\eqno(3.12)$$
 descends from the fivebrane charges (2.4), whereas the other two
 Noether charges,
$$ Q_{\b}= \int_{\S_3} {}^*_{4}J_{\b}\quad {\rm and}\quad
 Q_{\g}= \int_{\S_3} {}^*_{4}J_{\g}~,\eqno(3.13)$$ 
descend from the D2-brane (RR) charges. The existence of two charges is related
to the existence of two homology classes of a three-cycle $\cc$ \cite{bb}. The
 BPS brane charge quantization (sect.~2) implies that only a discrete subgroup
 of the continuous Peccei-Quinn type symmetries (3.6) and (3.7) is going to 
survive in full quantum theory after taking into account both membrane and 
fivebrane instanton corrections \cite{bb}. The discrete identifications of the 
UH scalars can be read off from eqs.~(3.6) and (3.7),
$$\eqalign{
 S~\sim~& S +in_{\h}+2(n_{\g}+in_{\b})C+n^2_{\g}+n^2_{\b}~,\cr
 C~\sim~& C+n_{\g}-in_{\b}~,\cr}\eqno(3.14)$$
where now all $n_{\h,\b,\g}$ are integers \cite{bb}. The transformations (3.14)
define a discrete non-abelian group $\cz$. The global (topological) structure 
of the UH moduli space $\cm$ is thus given by $Q/\cz$ \cite{bb}.

Since the type-IIA string loop corrections are invariant under the 
perturbative Peccei-Quinn type symmetries (3.6) and (3.7), it is natural to 
rewrite the UH metric in terms of the new quantities $(u,v,\f)$ defined by 
eq.~(A.14), which are all invariant under (3.6) and (3.7). When using the 
notation (A.14), it is not difficult to convince oneself that the unique 
quaternionic deformation of the classical UH action within the string 
perturbation theory is described by eq.~(2.18) indeed \cite{one}. Hence, the 
local UH metric (2.17) does not receive any perturbative quaternionic 
corrections modulo the UH field redefinitions (sect.~2).

\section{Hyper-K\"ahler versus quaternionic geometry}

The D-instanton corrections to the UH moduli space metric due to the wrapped
D2-branes were explicitly calculated in the hyper-K\"ahler limit by Ooguri
and Vafa \cite{ova}. In this limit the 4d, N=2 supergravity decouples 
\cite{bw},´ and the c-map applies \cite{fcg}.

In the absence of fivebrane corrections, the unbroken symmetry of the UH 
metric is given by $U_D(1)\times U_C(1)$ (sect.~3). In adapted coordinates
  with respect to the $U_D(1)$ isometry the UH metric can be written down in
 the LeBrun form \cite{lebrun}, 
$$ ds^2_{\rm K}\equiv g_{ab}d\f^ad\f^b=
W^{-1}(dt+\Theta_1)^2+W\left[ e^u(dx^2+dy^2)+d\o^2\right]~,\eqno(4.1)$$
where two potentials $W$ and $u$, and a one-form $\Theta_1$ have been
introduced. The metric Ansatz (4.1) is valid for {\it any} K\"ahler metric 
$g_{ab}$ in four real dimensions, $a,b=1,2,3,4$, with a Killing vector $K^a$ 
that preserves the K\"ahler structure. We use the adapted coordinates 
$\f^a=(t,x,y,\o)$ where $t$ is the coordinate along the trajectories of the 
Killing vector associated with $U_D(1)$, whereas $(x,y,\o)$ 
are the coordinates in the space of trajectories, $W^{-1}=g_{ab}K^aK^b\neq 0$.
Accordingly, no metric components are dependent upon $t$ in eq.~(4.1).

The K\"ahler condition on the metric (4.1) also implies a linear equation on 
$\Theta_1$ \cite{lebrun},
$$d\Theta_1=W_x\,dy\wedge d\o +W_y\,d\o\wedge dx+(We^u)_{\o}\,dx\wedge dy~.
\eqno(4.2)$$
In turn, this gives rise to the following integrability condition on $W$ 
\cite{lebrun}:
$$ W_{xx} + W_{yy}+(We^u)_{\o\o}=0~.\eqno(4.3)$$

The hyper-K\"ahler geometry implies, by definition, the existence of 
{\it three}
linearly independent K\"ahler structures $(J_k)\du{a}{b}$, $k=1,2,3$, 
which are covariantly constant, $\de_c(J_k)\du{a}{b}=0$, and obey a 
quaternionic algebra. Moreover, in four real dimensions, a hyper-K\"ahler
 metric necessarily has the {\it Anti-Self-Dual} (ASD) Riemann curvature, and 
vice versa \cite{besse}. As regards four-dimensional hyper-K\"ahler metrics, 
they are just K\"ahler and Ricci-flat, and vice versa \cite{ati}. 

If the Killing vector $K^a$ is triholomorphic (i.e. it is consistent with N=2
 supersymmetry), one may further restrict the metric (4.1) by taking $u=0$. 
The Riemann ASD condition then amounts to a {\it linear} system \cite{bf},
$$ \D W=0\quad {\rm and}\quad \vec{\de} W+\vec{\de}\times\vec{\Theta}=0~,
\eqno(4.4)$$
where $\D$ is the Laplace operator in three flat dimensions, 
$\D=4g^2_{\rm string}\pa_{z}\bar{\pa}_{\bar{z}}+\pa^2_{\o}$, and 
$z=g_{\rm string}(x+iy)$. The complex coordinate $z$ represents the RR-type
complex scalar, so that the unbroken $U_C(1)$ symmetry (3.8) implies that a 
solution to eq.~(4.4) may depend upon $z$ only via its absolute value 
$\abs{z}$. 

One may now think of $W$ as the electro-static potential for a collection of 
electric charges distributed in three dimensions near the axis $z=0$ with unit
 density in $\o$. The unique regular (outside the positions of charges) 
solution to this problem in the limit $g_{\rm string}\to 0$, while keeping 
$\abs{z}/g_{\rm string}$ finite, reads \cite{sh}  
$$ W =\fracmm{1}{4\p}\log\left(\fracmm{\m^2}{z\bar{z}}\right) +
\sum_{m=1}^{\infty} \fracmm{\cos(2\p m \o)}{\p}
K_0\left(\fracmm{2\p\abs{mz}}{g_{\rm string}}\right)~,\eqno(4.5)$$
where $K_0$ is the modified Bessel function. The solution (4.5) can be 
trusted  for large $\abs{z}$, where it amounts to the infinite 
 D-instanton/anti-instanton sum  \cite{ova}, 
$$\eqalign{
 W~=~&\fracmm{1}{4\p}\log \left( \fracmm{\m^2}{z\bar{z}}\right) +
\sum_{m=1}^{\infty} \exp \left(-\,\fracmm{2\p\abs{mz}}{g_{\rm string}}\right)
\cos(m\o)\cr
~& \times \sum_{n=0}^{\infty}\fracmm{\G(n+\fracm{1}{2})}{\sqrt{\p}n!
\G(-n+\fracm{1}{2})}\left(\fracmm{g_{\rm string}}{4\p\abs{mz}}
\right)^{n+\frac{1}{2}} ~~~~.\cr}\eqno(4.6)$$
The $\exp{(-1/g_{\rm string})}$ type dependence of the solution (4.6) 
apparently agrees with the general expectations \cite{w1} that it describes 
the D-instantons indeed.

If merely the vanishing {\it scalar} curvature of the K\"ahler 
metric (4.1) or a non-triholomorphic isometry of the hyper-K\"ahler metric 
were required, we would end up with the non-linear equation \cite{lebrun,bf}
$$ u_{xx} + u_{yy}+(e^u)_{\o\o}=0~.\eqno(4.7)$$
The equation (4.7) is known as the $SU(\infty)$ Toda field equation 
\cite{ward} since it appears in the large-$N$ limit of the standard 
(two-dimensional) Toda system for $SU(N)$. See, e.g., ref.~\cite{lez} for more
 about the Toda equation (4.7),  and ref.~\cite{nlsm} for more about the
four-dimensional hyper-K\"ahler NLSM. The Einstein-K\"ahler deformations of the
Bergmann metric of $SU(2,1)/U(2)$  in eq.~(A.1) were investigated in 
refs.~\cite{fef,bab}.

\section{D-instantons and quaternionic UH metric}

A quaternionic manifold admits three independent {\it almost} complex 
structures $(\tilde{J}_k)\du{a}{b}$, which are, however, {\it not} covariantly 
constant but satisfy  
$\de_a(\tilde{J}_k)\du{b}{c}= (T_a)\du{k}{n}(\tilde{J}_n)\du{b}{c}$, where
 $(T_a)\du{k}{n}$ is the NLSM torsion \cite{besse}. This torsion is induced
by 4d, N=2 supergravity because the quaternionic condition on the 
hypermultiplet NLSM target space metric is the direct consequence of 
{\it local} N=2 supersymmetry in four spacetime dimensions \cite{bw}. 
As regards four-dimensional quaternionic manifolds (relevant for UH), they all
 have {\it Einstein-Weyl\/} geometry of {\it negative\/} scalar curvature 
\cite{bw,besse}, i.e.
$$ W^-_{abcd}=0~,\qquad R_{ab}=\fracm{\L}{2}g_{ab}~,\eqno(5.1)$$  
where  $W_{abcd}$ is the Weyl tensor and  $R_{ab}$ is the Ricci tensor for 
the metric $g_{ab}$. The overall coupling constant of the 4d NLSM has the same
dimension as $\k_4^2$, while in the N=2 locally supersymmetric NLSM these 
coupling constants are proportional to each other with the dimensionless
 coefficient $\L<0$ \cite{bw}. We take $\k^2_4=1$ for simplicity. 

Since the quaternionic and hyper-K\"ahler conditions are not compatible, the 
canonical form (4.1) should be revised.~\footnote{A generic Einstein-Weyl 
manifold does not have a K\"ahler structure.} Nevertheless, the exact 
quaternionic metric is governed  by {\it the same\/} three-dimensional 
Toda equation (4.7) \cite{din}. Indeed, when using another (Tod) {\it Ansatz} 
\cite{tod} 
$$ ds^2_{\rm Q}= \fracmm{P}{\o^2}\left[ e^u (dx^2+dy^2)+d\o^2\right]
+\fracmm{1}{P\o^2}(dt +\Theta_1)^2 \eqno(5.2)$$
for a quaternionic metric with an abelian isometry, 
it is straightforward to prove that the restrictions (5.1) on the metric (5.2)
 {\it precisely\/} amount to eq.~(4.7) on the potential 
$u=u(x,y,\o)$, while $P$ is given by \cite{tod}
$$ P= \fracmm{1}{2\L}\left(\o u_{\o}-2\right),\eqno(5.3)$$
whereas the one-form $\Theta_1$ obeys the linear equation \cite{tod}
$$ d\Theta_1= -P_x\,dy\wedge d\o-P_y\,d\o\wedge dx-e^u(P_{\o}+\fracm{2}{\o}P
+\fracm{2\L}{\o}P^2)dx\wedge dy~.\eqno(5.4)$$ 

The limit $\L\to 0$, where 4d, N=2 supergravity decouples, should be taken 
with care. After rescaling $u\to \L u$ in eq.~(4.7) we get
$$ u_{xx} + u_{yy}+\fracmm{1}{\L}(e^{\L u})_{\o\o}=0~.\eqno(5.5)$$
This equation gives rise to the 3d Laplace (linear!) equation when  $\L\to 0$,
as expected. We can, therefore, conclude that the non-linear Toda equation 
(4.7) substitutes the linear Laplace equation (4.4) in the presence of 4d,
N=2 supergravity. The `cosmological' constant $\L$ can be considered as the 
 most relevant parameter of the deformation that converts a given UH
hyper-K\"ahler metric into the `gravitationally dressed' quaternionic UH 
metric via eq.~(5.2), based on {\it the same} solution to the Toda equation 
(4.7) \cite{din}.

In terms of the complex coordinate $\z=x+iy$, the 3d Toda equation (4.7) takes 
the form
$$ 4u_{\z\bar{\z}}+(e^u)_{\o\o}=0~.\eqno(5.6)$$
It is not difficult to check that this equation is invariant under  
{\it holomorphic} transformations of $\z$,
$$\z\to \hat{\z}=f(\z)~,\eqno(5.7)$$
with an arbitrary function $f(\z)$, provided that it is accompanied by the 
shift of the Toda potential,
$$ u \to \hat{u} = u -\log (f') -\log (\bar{f}')~,\eqno(5.8)$$
where the prime means differentiation with respect to $\z$ or $\bar{\z}$, 
respectively. The transformations (5.7) can be interpreted as the residual
diffeomorphisms in the NLSM target space of the universal hypermultiplet, 
which keep invariant the quaternionic Ansatz (5.2) under the compensating 
`Toda gauge transformations' (5.8). 

To make contact with particular four-dimensional hyper-K\"ahler geometries 
(with isometries) \cite{nlsm}, it is natural to search for {\it separable} 
exact solutions to the Toda equation, having the form
$$ u (\z,\bar{\z},\o) = F(\z,\bar{\z}) + G(\o)~.\eqno(5.9)$$
Equation (4.7) now reduces to two separate equations,
$$ F_{\z\bar{\z}}+\fracm{c^2}{2}e^F=0 \eqno(5.10)$$
and
$$ \pa^2_{\o}e^G=2c^2~,\eqno(5.11)$$
where $c^2$ is a separation constant. After taking into account the 
positivity of $e^G$, the  general solution to eq.~(5.11) reads
$$ e^G = c^2(\o^2 + 2\o b\cos\a + b^2)~,\eqno(5.12)$$
where $b$ and $\a$ are arbitrary real integration constants.

Equation (5.10) is the 2d {\it Liouville} equation that is well known in 2d 
quantum gravity \cite{cftbook}. Its general solution reads 
$$ e^F = \fracmm{4\abs{f'}^2}{(1+c^2\abs{f}{}^2)^2} \eqno(5.13)$$
in terms of arbitrary holomorphic function $f(\z)$. The ambiguity associated
with this function is, however, precisely compensated by the Toda gauge
transformation (5.8), so that we have the right to choose $f(\z)=\z$ 
in eq.~(5.13). This yields the following regular exact solution to the 3d Toda 
equation:
$$ e^u= \fracmm{4c^2(\o^2+2\o b\cos\a +b^2)}{(1+c^2\abs{\z}{}^2)^2}~~~.
\eqno(5.14)$$
It is obvious now that the konstant $c^2$ is positive indeed. It also follows 
from eqs.~(5.2), (5.3) and (5.14) that any separable exact solution to the 
quaternionic UH metric possesses the rigid $U_C(1)$ duality symmetry with 
respect to the duality rotations $\z\to e^{i\a}\z$ of the complex RR-field 
$\z$.

Though the quaternionic NLSM metric defined by eqs.~(5.2), (5.3), (5.4) and 
(5.14) is apparently different from the classical UH metric (2.17), these 
metrics are nevertheless equivalent in the classical region of the UH moduli 
space where all quantum corrections are suppressed. The classical approximation
corresponds to the conformal limit $\o\to\infty$ and $\abs{\z}\to\infty$, 
while keeping the ratio $\abs{\z}^2/\o$ finite. Then one easily finds that 
$P\to -\L^{-1}=const.>0$, whereas the metric (5.2) takes the form
$$ ds^2=\fracmm{1}{\l^2}\left(\abs{dC}^2+d\l^2\right)
+\fracmm{1}{\l^4}(dD+\Theta)^2~,\eqno(5.15)$$
in terms of the new variables $C=1/\z$ and $\l^2=\o$, after a few rescalings.
The metric (5.15) reduces to that of eq.~(2.17) when using $\l^{-2}=e^{2\f}$. 
Another interesting limit is $\o\to 0$ and $\abs{\z}\to\infty$, where one gets
a conformally flat metric $(AdS_4)$.

Based on the fact that both hyper-K\"ahler and quaternionic metrics under 
consideration are governed by  the same Toda equation, a natural 
mechanism~\footnote{The different mechanism, which associates the
quaternionic metric in $4(n+1)$ real dimensions \newline ${~~~~~}$ to a 
given (special) hyper-K\"ahler metric in $4n$ real dimensions, was proposed in 
ref.~\cite{witk}.} of generating the quaternionic metrics from known 
hyper-K\"ahler metrics in the same (four) dimensions arises: first, 
one deduces a solution to the Toda equation (4.7) from a given 
four-dimensional hyper-K\"ahler metric having a non-triholomorphic or 
rotational isometry, by rewriting it to the form (4.1), and then one inserts 
the obtained exact solution into the quaternionic Ansatz (5.2) to deduce the 
corresponding  quaternionic metric with the same isometry. Being applied to 
the D-instantons, this mechanism results in their dressing with respect to 4d,
N=2 supergravity background \cite{din}.

The $SU(\infty)$ Toda equation is known to be notoriously difficult to solve,
while a very few its exact solutons are known \cite{lez}. Nevertheless, the 
proposed connection to the hyper-K\"ahler metrics can be used as the powerful 
vehicle for generating exact solutions to eq.~(4.7). It is worth mentioning 
that eq.~(4.3) follows from eq.~(4.7) after a substitution 
$$ W=\pa_{\o}u~,\eqno(5.16)$$
while eq.~(5.2) is solved by
$$ \Theta_1= \mp\pa_y u (dx)  \pm\pa_x u (dy)~.\eqno(5.17)$$
This is known as the Toda frame for a hyper-K\"ahler metric 
\cite{bf,bak,bak2}. It is not difficult to verify that the separable solution 
(5.14) is generated from the Eguchi-Hanson (hyper-K\"ahler) metric along these
 lines \cite{bak,nlsm}. Another highly non-trivial solution to the Toda 
equation (4.17) follows from the Atiyah-Hitchin  (hyper-K\"ahler) metric 
\cite{ati,bak2}. The transform to the Toda frame for the Atiyah-Hitchin metric
is known  \cite{ol},
$$ y+{\rm i}x = K(k)\sqrt{1+{k'}^2\sinh^2\n}
\left(\cos\vq +\fracmm{\tanh\n}{K(k)}
\int^{\p/2}_{0} {\rm d}\g\,
\fracmm{\sqrt{1-k^2\sin^2\g}}{1-k^2\tanh^2\n\sin^2\g}\right)$$
$$ \o= \fracmm{1}{8}K^2(k)\left( k^2\sin^2\vq+{k'}^2(1+\sin^2\vq\sin^2\j)-
\fracmm{2E(k)}{K(k)}\right),\eqno(5.18)$$
where $(\vq,\j,\vf;k)$ are the new coordinates (in four dimensions). The
parameter $k$ plays the role of modulus here, $0<k<1$, while 
$k'=\sqrt{1-k^2}$ is called the complementary modulus. The remaining
 definitions are
$$\n\equiv \log\left(\tan\fracmm{\vq}{2}\right)+i\j~,\quad
\t = 2\left(\vf +{\rm arg}(1+{k'}^2\sinh^2\n)\right)~,\eqno(5.19)$$
in terms of the standard complete elliptic integrals of the first and second 
kind, $K(k)$ and $E(k)$, respectively. The solution to the Toda equation (4.7)
now reads \cite{bak,nlsm}
$$ {\rm e}^{u}=\fracmm{1}{16}K^2(k)\sin^2\vq\abs{1+{k'}^2\sinh^2\n}~.
\eqno(5.20)$$
The physical significance of the related quaternionic metric solution is 
apparent in the perturbative region $k\to 1$, where \cite{k2} 
$$  k'\propto e^{-S_{\rm inst.}}~,\quad{\rm and}\quad 
S_{\rm inst.}\to +\infty~.\eqno(5.21)$$
In this limit the Atiyah-Hitchin metric is exponentially close to the Taub-NUT
 metric \cite{ati}, while the exponentially small corrections can be 
interpreted as the (mixed) D instantons and anti-instantons. The D-instanton 
action $S_{\rm inst.}$ represents here the volume of the corresponding 
suppersymmetric three-cycle $\cc$, on which the D-brane is wrapped \cite{gs}.
It is worth mentioning that the same exact solution also describes the 
hypermultiplet moduli space metric in the 3d, N=4 supersymmetric Yang-Mills 
theory with the $SU(2)$ gauge group, which was obtained via the c-map 
in ref.~\cite{sw}.

\section{Fivebrane instantons and Tod-Hitchin metric}

As was demonstrated in ref.~\cite{gs}, the BPS condition on the fivebrane 
instanton solution with the vanishing charge $Q_{\h}$ of eq.~(3.12) defines a 
gradient flow in the hypermultiplet moduli space. The flow implies the $SU(2)$
isometry of the UH metric since the non-degenerate action of this isometry in 
the four-dimensional UH moduli space gives rise to the well defined 
three-dimensional orbits which can be parametrized by the `radial'  coordinate
to be identified with the flow parameter. In the case of the classical UH 
moduli space $SU(2,1)/SU(2)\times U(1)$ the radial coordinate $(r)$ is defined 
by eq.~(A.5), while its relation to the complex $(S,C)$ coordinates is given by
$$ r^2=\fracmm{\abs{1-S}^2+4\abs{C}^2}{\abs{1+S}^2}~~~,\eqno(6.1)$$
where we have used eqs.~(3.5) and (A.5). It is worth mentioning that the  
non-abelian $SU(2)$ symmetry includes the abelian duality rotations (3.8).
However, it does not imply the preservation of the rest of the 
$SU(2,1)/SU(2)\times U(1)$ symmetries (other than $SU(2)$) including the 
Peccei-Quinn type symmetries (3.6) and (3.7). Hence, the $SU(2)$-invariant 
deformations of the classical UH metric, subject to the quaternionic 
constraints (5.1), should describe the zero-charge fivebrane instanton 
corrections.

The action of a fivebrane instanton with the vanishing charge may merely 
depend upon the complex structure modulus at infinity \cite{gs}. Therefore, we
expect the corresponding UH moduli space metric be merely dependent upon the 
complex structure on the boundary of the coset $SU(2,1)/SU(2)\times U(1)$. The
 conformal boundary metric for the coset $SU(2,1)/SU(2)\times U(1)$ is known 
to be degenerate, i.e. it possess a zero eigenvalue. This happens because the 
conformal structure, associated with the Bergmann metric in the form (A.6) 
inside the unit ball in ${\bf C}^2$, does not extend across the boundary since
 the coefficient at $\s^2_2$ in eq.~(A.6) decays {\it faster\/} than the 
coefficients at $\s^2_1$ and  $\s^2_3$. However, the conformal structure 
survives  in the {\it two-dimensional} (2d) subspace annihilated by $\s_2$. 
The 2d complex structure has a single real parameter -- the central charge 
\cite{cftbook} -- which should appear in the UH metric. The exact metric
solutions, given below in this section, confirm these expectations.

Let's consider a generic $SU(2)$-invariant metric in four Euclidean dimensions.
 In the Bianchi IX formalism, where the $SU(2)$ symmetry is manifest, the 
general {\it Ansatz} for such metrics reads (see Appendix A for our notation)
$$ ds^2=w_1w_2w_3dt^2+\fracmm{w_2w_3}{w_1}\s^2_1+\fracmm{w_3w_1}{w_2}\s^2_2+  
\fracmm{w_1w_2}{w_3}\s^2_3~,\eqno(6.2)$$
in terms of the (left)-invariant one-forms $\s_i$ defined by eq.~(A.7), and the
radial coordinate $t$. The metric (6.2) is dependent upon three functions
$w_i(t)$, $i=1,2,3$. The most general $SU(2)$-invariant metric is given by a
quadratic form with respect to $\s_i$. However, it can always be chosen in the
 diagonal form, as in eq.~(6.2), without loss of generality. 

The quaternionic constraints on the metric amount to the ASD-Weyl equation
and the Einstein equation --- see eq.~(5.1). The exact $SU(2)$-invariant 
solutions to these equations were found by Tod \cite{tod3} and Hitchin 
\cite{hit}. The main results of refs.~\cite{tod3,hit} are briefly described 
below.

Being applied to the metric (6.2), the ASD Weyl condition gives rise to a
system of {\it Ordinary Differential Equations} (ODE) \cite{tod3},
$$\eqalign{
 \dt{A}_1~=~&-A_2A_3 +A_1(A_2+A_3) ~,\cr
 \dt{A}_2~=~&-A_3A_1 +A_2(A_3+A_1) ~,\cr
 \dt{A}_3~=~&-A_1A_2 +A_3(A_1+A_2) ~,\cr}\eqno(6.3)$$
where the dots denote differentiation with respect to $t$, and the functions 
$A_i(t)$ are defined by the auxiliary ODE system,
$$\eqalign{
   \dt{w}_1~=~&-w_2w_3 +w_1(A_2+A_3)~,\cr
   \dt{w}_2~=~&-w_3w_1 +w_2(A_3+A_1)~,\cr
   \dt{w}_3~=~&-w_1w_2 +w_3(A_1+A_2)~.\cr}\eqno(6.4)$$

The ODE system (6.3) is known in the mathematical literature as the 
classical Halphen system \cite{halphen}. The Bergmann metric (A.6) is the
simplest solution to eqs.~(6.3) and (6.4) with $A_i=0$ for all $i=1,2,3$. This 
follows from comparison of eqs.~(6.3), (6.4) and (A.6). Note that despite of 
the fact that all $A_i=0$, the Bergmann metric (A.6) is, nevertheless, 
non-trivial (or non-flat) since eq.~(A.6) is still a non-trivial solution to 
eq.~(6.4).

Given a metric solution to the ASD Weyl equations, the Einstein equation of 
eq.~(5.1) can be easily satisfied after proper Weyl rescaling of the ASD Weyl 
metric, because any local Weyl transformation does not affect the vanishing 
Weyl tensor. Having obtained an explicit solution to the Halphen system (6.3),
 it may be substituted into the ODE system (6.4). To solve eq.~(6.4), it is 
convenient to change variables as \cite{tod3}
$$ \eqalign{
 w_1~=~&\fracmm{\O_1\dt{x}}{\sqrt{x(1-x)}}~,\cr
 w_2~=~&\fracmm{\O_2\dt{x}}{\sqrt{x^2(1-x)}}~,\cr
 w_3~=~&\fracmm{\O_3\dt{x}}{\sqrt{x(1-x)^2}}~~,\cr}\eqno(6.5)$$
where the new variables $\O_i(x)$, $i=1,2,3$, are constrained by an algebraic
condition,
$$\O_2^2+\O_3^2-\O^2_1=\fracm{1}{4}~~. \eqno(6.6) $$
The algebraic relation (6.6) reduces the number of the newly introduced 
functions in eq.~(6.5) from four to three, as it should. In fact, eq.~(6.6) is
also dictated by the quaternionic nature of the metric \cite{tod3,hit}.

In terms of the new variables (6.5), the ODE system (6.4) takes the form 
\cite{tod3,hit}
$$\eqalign{
 \O_1'~=~&-\fracmm{\O_2\O_3}{x(1-x)}~,\cr
 \O_2'~=~&-\fracmm{\O_3\O_1}{x}~,\cr
 \O_3'~=~&-\fracmm{\O_1\O_2}{1-x}~,\cr}\eqno(6.7)$$
where the primes denote differentiation with respect to $x$. It is not
difficult to verify that the algebaric constraint (6.6) is preserved under 
the flow (6.7), so that the highly non-linear transformation (6.5) is fully 
consistent. In terms of the new variables $(x,\O_i)$, the Einstein condition
of eq.~(5.1) on the metric (6.2) in the form
$$ ds^2=e^{2u}\left[ \fracmm{dx^2}{x(1-x)}+\fracmm{\s^2_1}{\O^2_1}
+\fracmm{(1-x)\s^2_2}{\O^2_2}+\fracmm{x\s^2_3}{\O^2_3}\right] \eqno(6.8)$$
amounts to the algebraic relation \cite{tod3}
$$ 96\k^2e^{2u}=\fracmm{8x\O^2_1\O^2_2\O^2_3+2\O_1\O_2\O_3(x(\O_1^2+\O_2^2)-
(1-4\O^2_3)(\O^2_2-(1-x)\O ^2_1))}{(x\O_1\O_2+2\O_3(\O_2^2-(1-x)\O^2_1))^2}
\eqno(6.9)$$
which yields the Weyl factor $u(x)$ in terms of the functions $\O_i(x)$.

The Halphen system (6.3) has a long history \cite{abc}. Perhaps, its most
natural (manifestly integrable) derivation is provided via a reduction of 
the $SL(2,{\bf C})$ anti-self-dual Yang-Mills equations from four Euclidean 
dimensions to one  \cite{int}. A classification of all possible reductions is 
known in terms of the so-called {\it Painlev\'e} groups that give rise to six 
different types of integrable Painlev\'e equations \cite{int}. It remains to 
identify those of them that lay behind the ASD-Weyl (or quaternionic-K\"ahler)
geometry with the $SU(2)$ symmetry. There are only two natural (or nilpotent, 
in the terminology of ref.~\cite{int}) types (III and VI) that give rise to a
 single non-linear integrable equation. In the geometrical terms, it is the 
Painlev\'e III equation that lays behind the four-dimensional K\"ahler spaces 
with vanishing scalar curvature \cite{pp}, whereas the Painlev\'e VI equation 
is known to be behind the ASD-Weyl geometries having the $SU(2)$ symmetry 
\cite{tod3,hit,m3}. A generic Painlev\'e VI equation has four real parameters 
\cite{int}, but they are all fixed by the quaternionic property 
\cite{tod3,hit}, as in eq.~(6.6). This means that the quaternionic metrics with
the $SU(2)$ symmetry are all governed by the particular Painlev\'e VI equation:
$$\eqalign{
y''~=~&\fracmm{1}{2}\left( \fracmm{1}{y} 
+\fracmm{1}{y-1}+\fracmm{1}{y-x}\right)
(y')^2- \left( \fracmm{1}{x} +\fracmm{1}{x-1}+\fracmm{1}{y-x}\right)y'\cr
&~  +\fracmm{y(y-1)(y-x)}{x^2(x-1)^2}\left[
\fracmm{1}{8} -\fracmm{x}{8y^2}+\fracmm{x-1}{8(y-1)^2}
+\fracmm{3x(x-1)}{8(y-x)^2}\right] ~,\cr}\eqno(6.10)$$
where $y=y(x)$, and the primes denote differentiation with respect to $x$.

The equivalence between eqs.~(6.3) and (6.10) via eqs.~(6.4) and (6.5) 
is known to mathematicians \cite{tod3,hit,m3}. Explicitly, in the Einstein 
case, it is given by the relations
$$\eqalign{
\O^2_1~=~&\fracmm{(y-x)^2y(y-1)}{x(1-x)}\left(v-\fracmm{1}{2(y-1)}\right)\left(
v-\fracmm{1}{2y}\right)~,\cr
\O^2_2~=~&\fracmm{(y-x)y^2(y-1)}{x}\left(v-\fracmm{1}{2(y-x)}\right)\left(
v-\fracmm{1}{2(y-1)}\right)~,\cr
\O^2_3~=~&\fracmm{(y-x)y(y-1)^2}{(1-x)}\left(v-\fracmm{1}{2y}\right)\left(
v-\fracmm{1}{2(y-x)}\right)~,\cr}\eqno(6.11)$$
where the auxiliary variable $v$ is defined by the equation
$$ y{}'=\fracmm{y(y-1)(y-x)}{x(x-1)}\left(2v-\fracmm{1}{2y}-\fracmm{1}{2(y-1)}
+\fracmm{1}{2(y-x)}\right)~.\eqno(6.12)$$

An exact solution to the Painlev\'e VI equation (6.10), which leads to a 
{\it regular} (and complete) quaternionic metric (6.2), is unique \cite{hit}.
The Hitchin solution \cite{hit} can be expressed in terms of the standard 
theta-functions $\vq_{\a}(z|\t)$, where $\a=1,2,3,4$.~\footnote{We use the 
standard definitions and notation for the theta functions \cite{theta} --- 
see Appendix B.} 

In order to write down the Hitchin solution to eq.~(6.10), 
the theta-function arguments should be related, $z=\fracm{1}{2}(\t -k)$, 
where $k$ is an arbitrary (real and positive) parameter. The variable $\t$ is
related to the variable $x$ of eq.~(6.10) via
$$x=\vq^4_3(0)/\vq^4_4(0)~,\eqno(6.13)$$ 
where the value of the variable $z$ is explicitly indicated, as usual.
 One finds \cite{dub,hit}
$$\eqalign{
y(x)~=~& \fracmm{\vq_1'''(0)}{3\p^2\vq^4_4(0)\vq_1'(0)}+\fracmm{1}{3}\left[
1+\fracmm{\vq_3^4(0)}{\vq^4_4(0)}\right] \cr
 &~ +\fracmm{\vq_1'''(z)\vq_1(z)-2\vq_1''(z)\vq_1'(z)+
2\p i(\vq_1''(z)\vq_1(z)-\vq_1'{}^2(z))}{2\p^2\vq_4^4(0)\vq_1(z)(\vq_1'(z)+
\p i\vq_1(z))}~~.\cr}\eqno(6.14)$$ 

The parameter $k>0$ describes the monodromy of the solution (6.14) around 
its essential singularities (branch points) $x=0,1,\infty$. This (non-abelian)
 monodromy is generated by the matrices (with the purely imaginary eigenvalues
 $\pm i$) \cite{hit}
$$ M_1=\left( \begin{array}{cc} 0 & i \\ i & 0\end{array} \right)~,\quad
  M_2=\left( \begin{array}{cc} 0 & i^{1-k} \\ 
i^{1+k} & 0\end{array} \right)~,\quad
M_3=\left( \begin{array}{cc} 0 & i^{-k} \\ 
-i^{k} & 0\end{array} \right)~.\eqno(6.15)$$

Another explicit (equivalent) form of the Hitchin exact solution to the metric 
coefficients $w_i$ in eqs.~(6.2) and (6.4) was derived in ref.~\cite{bk}, 
in terms of the theta functions with characteristics, by the use of the 
fundamental Schlesinger system and the isomonodromic deformation techniques. 

The function (6.14) is meromorphic outside its essential singularities at 
$x=0,1,\infty$, while is also has simple poles at 
$\bar{x}_1,\bar{x}_2,\ldots$, where $\bar{x}_n\in (x_n,x_{n+1})$ and 
$x_n=x(ik/(2n-1))$ for each positive integer $n$. Accordingly, the metric is 
well-defined (complete) for $x\in (\bar{x}_n,x_{n+1}]$, i.e. in the unit ball 
with the origin at $x=x_{n+1}$ and the boundary at $x=\bar{x}_n$ \cite{hit}.
Near the boundary the Tod-Hitchin metric (6.2) has the asymptotical behaviour 
$$\eqalign{
 ds^2~=~&\fracmm{dx^2}{(1-x)^2}+\fracmm{4}{(1-x)\cosh^2(\p k/2)}\s^2_1+
\fracmm{16}{(1-x)^2\sinh^2(\p k/2)\cosh^2(\p k/2)}\s^2_2\cr
&~ + \fracmm{4}{(1-x)\sinh^2(\p k/2)}\s^2_3~~+ ~~{\rm regular~terms}~.\cr}
\eqno(6.16)$$

As is clear from eq.~(6.16), the coefficient at $\s^2_2$ vanishes faster than
the coefficients at $\s^2_1$ and $\s^2_3$ when approaching the boundary,
$x\to 1^-$, similarly to eq.~(A.6). On the two-dimensional boundary 
annihilated by $\s_2$ one has the natural conformal structure
$$ \sinh^2(\p k/2)\s^2_1 +\cosh^2(\p k/2)\s^2_3~. \eqno(6.17)$$
The only relevant parameter $\tanh^2(\p k/2)$ in eq.~(6.17) represents the 
central charge (or the conformal anomaly) on the boundary. This result is 
apparently consistent with (i) the fact that the type-IIA fivebranes can be 
described in terms of an exact (super)conformal field theory \cite{cal}, 
(ii) the Zamolodchikov $c$-theorem \cite{zam2}, and (iii) the holographic 
principle \cite{last}.

The constraints (5.1) do not seem to imply any quantization condition on the 
monodromy parameter $k$ since the regular metric solutions exist for any $k>0$.
 The central charge (or the critical exponent $k$) is quantized in solvable 
2d, N=2 superconformal field theories (the minimal N=2 superconformal models) 
which are associated with {\it compact} (simply-laced) Lie groups 
\cite{cftbook}. The absence of central charge quantization in our case may be 
related to the {\it negative} scalar curvature of the metrics. The 
Einstein-Weyl metrics of the {\it positive} scalar curvature take the similar 
form given by eqs.~(6.2) or (6.8), while they are known to be related to the 
so-called Poncelet $n$-polygons that give rise to the quantization condition 
$k=2/n$, where $n\in {\bf Z}$ \cite{hit2}. It is thus the {\it non-compact} 
nature of the Lie group $SU(2,1)$ that is responsible for the absence of the
central charge quantization in the boundary conformal field theory. 

\section{Conclusion}

The universal hypermultiplet gives us the unique opportunity to learn more 
about the non-perturbative quantum corrections in string/M-theory via better 
understanding of the exact quaternionic geometry governing the hypermultiplet 
LEEA (or NLSM) in the 4d, N=2 supergravity background.  

The D-instanton corrections to the classical UH moduli space metric are 
calculable due to the residual $U_D(1)\times U_C(1)$ symmetry (sects.~4 and 5).
 No such corrections arise when all the other fields (except UH) are turned 
off. The fivebrane instanton corrections are calculable at vanishing fivebrane
 charges (sect.~6). In a generic case with non-vanishing charges, there may 
be also contributions from the membranes ending on fivebranes wrapped on a
CY space \cite{gs}. Then only $U_C(1)$ isometry may be left. However, the 
quaternionic Ansatz (5.2) still applies, this time with respect to the 
$U_C(1)$ isometry. The absence of any other symmetry means that generic 
instanton corrections are described by {\it non-separable} exact solutions to
the three-dimensional Toda equation (4.7). This observation is consistent with
another observation that the instanton action is not a sum of membrane and 
fivebrane contributions \cite{gs}.

More general (than Atiyah-Hitchin) regular hyper-K\"ahler four-manifolds with 
a rotational isometry are known \cite{dan}, though in the rather implicit form
(as the algebraic curves). It is also known that the Toda equation (4.7) can 
be reduced in a highly non-trivial way to the Painlev\'e equations \cite{tod2}.
It is not clear to us how to extract the specific information from this 
knowledge, which would be relevant for the problem mentioned in the title.

\section*{Acknowledgements}

I am grateful to Klaus Behrndt, Vicente Cortes, Evgeny Ivanov, Olaf 
Lechtenfeld, Werner Nahm, Hermann Nicolai, Paul Tod, Galliano Valent, Stefan 
Vandoren and Bernard de Wit for discussions. This paper is based on the 
invited talks given by the author at the Institute for Theoretical Physics in 
Hannover and the Max-Planck-Institute for Mathematics in Bonn, within the 
national (DFG) programm `String Theory'.
\vglue.2in

\section*{Appendix A: coset space $SU(2,1)/SU(2)\times U(1)$}

The homogeneous symmetric space $SU(2,1)/SU(2)\times U(1)$ is topologically
equivalent to the open ball in ${\bf C}^2$ with the Bergmann metric 
\cite{besse}
$$ ds^2= \fracmm{dz_1d\bar{z}_1+dz_2d\bar{z}_2}{1-\abs{z_1}{}^2-\abs{z_2}{}^2}
+ \fracmm{(\bar{z}_1dz_1+\bar{z}_2dz_2)(z_1d\bar{z}_1+z_2d\bar{z}_2)}{(1-
\abs{z_1}{}^2-\abs{z_2}{}^2)^2}~~~,\eqno(A.1)$$
where $\abs{z_1}^2+\abs{z_2}^2< 1$ and $\abs{z}{}^2=z\bar{z}$. 
The metric (A.1) is K\"ahler,
$$ ds^2=(\pa_a\bar{\pa}_b K)dz_ad\bar{z}_b=e^Kdz_ad\bar{z}_a+
e^{2K}(\bar{z}_adz_a)(z_bd\bar{z}_b)~,\qquad a,b=1,2,\eqno(A.2)$$
with the K\"ahler potential 
$$ K = - \log(1-\abs{z_1}{}^2-\abs{z_2}{}^2)~.\eqno(A.3)$$
The K\"ahler potential is defined modulo the K\"ahler gauge 
transformations,
$$ K(z,\bar{z})\to \tilde{K}(z,\bar{z}) = K(z,\bar{z})
 + f(z) +\bar{f}(\bar{z})~, \eqno(A.4)$$
with an arbitrary {\it holomorphic} function $f(z_1,z_2)$. 

Equation (A.3) clearly shows that the Bergmann metric is dual to the 
Fubini-Study metric on the compact complex projective space 
${\bf CP}^2=SU(3)/SU(2)\times U(1)$ \cite{cp2}. The homogeneous space 
${\bf CP}^2$ is symmetric, while it is also an Einstein space of positive 
scalar curvature with the (anti)self-dual Weyl tensor.~\footnote{The Weyl 
tensor is the traceless part of the Riemann curvature.} The non-compact coset 
space $SU(2,1)/U(2)$ is, therefore, also an Einstein space (though of negative
 scalar curvature), with the (anti)self-dual Weyl tensor. In other words, the 
coset space $SU(2,1)/U(2)$ is an Einstein-Weyl (or a self-dual Einstein) 
space. In four dimensions, the Einstein-Weyl spaces are called 
{\it quaternionic} by definition \cite{besse}. 

After the coordinate change
$$ z_1=r\cos\fracm{\q}{2}e^{i(\vf+\j)/2}~,\quad
 z_2=r\sin\fracm{\q}{2}e^{-i(\vf-\j)/2}~,\eqno(A.5)$$
the metric (A.1) can be rewritten to the diagonal form in the Bianchi IX 
formalism with manifest $SU(2)$ symmetry,
$$ ds^2=\fracmm{dr^2}{(1-r^2)^2} +\fracmm{r^2\s^2_2}{(1-r^2)^2}+
\fracmm{r^2}{(1-r^2)}(\s_1^2+\s_3^2)~,\eqno(A.6)$$
where we have introduced the $su(2)$ (left)-invariant one-forms
$$ \eqalign{
\s_1= & -\fracm{1}{2} \left( \sin\j\sin\q d\vf +\cos\j d\q\right)~,\cr
\s_2= & \fracm{1}{2} \left( d\j +\cos\q d\vf\right)\,\cr
\s_3= & \fracm{1}{2} \left( \sin\j d\q  -\cos\j\sin\q d\vf\right)~,\cr}
\eqno(A.7)$$
in terms of four real coordinates $0\leq r<1$,  $0\leq \q < \p$,  
$0\leq \vf< 2\p$ and $0\leq \j< 4\p$. The one-forms (A.6) obey the relations
$$ \s_i\wedge \s_j =\fracm{1}{2}\ve_{ijk}d\s_k~,\quad i,j,k=1,2,3~.\eqno(A.8)$$

Another useful parametrization of the coset $SU(2,1)/U(2)$ arises after the
following change of variables: 
$$ z_1= \fracmm{1-S}{1+S}~,\quad  z_2= \fracmm{2C}{1+S}~~~.\eqno(A.9)$$
The K\"ahler potential in terms of the new complex variables $(S,C)$ reads
$$ \ck =-\log \left( S+\bar{S} - 2C\bar{C}\right)~,\eqno(A.10)$$
with the  K\"ahler metric
$$ ds^2= e^{2\ck}\left(dSd\bar{S} -2CdSd\bar{C}-2\bar{C}d\bar{S}dC+
2(S+\bar{S})dCd\bar{C}\right)~.\eqno(A.11)$$
This form of the metric was used by Ferrara and Sabharwal \cite{fsh} in their
analysis of the type-II superstring vacua on Calabi-Yau spaces, in terms of the
complex RR-type scalar $C$ and the complex scalar
$$ S = e^{-2\f}+2iD + C\bar{C}~,\eqno(A.12)$$
where $\f$ stands for the dilaton field and $D$ stands for the axion field in 
four spacetime dimensions. The metric (A.11) can also be rewritten to the form
$$ ds^2=\abs{u}{}^2+\abs{v}{}^2~, \eqno(A.13)$$
where 
$$ u \equiv e^{\f}dC\quad {\rm and} \quad v\equiv 
e^{2\f}\left(\fracm{1}{2}dS-\bar{C}dC\right)~,\eqno(A.14)$$
by using the relations
$$\f=-\fracm{1}{2}\ln\left[( S+\bar{S}-2C\bar{C})/2\right]\eqno(A.15)$$
and 
$$ e^{\ck} =\fracm{1}{2}e^{2\f}~,\eqno(A.16)$$  
in accordance with eqs.~(A.10) and (A.12). The notation (A.13) and (A.14) 
was used by Strominger \cite{one} in his investigation of the one-loop string
corrections to the universal hypermultiplet. Of course, the metric (A.13) 
is not flat, since the one-forms $u$ and $v$ of eq.~(A.14) are not exact. 
Here are some useful identities \cite{one}
$$ \eqalign{
du~=~& \fracm{1}{2}u\wedge (v+\bar{v})~,\cr
dv~=~& v\wedge \bar{v} + u\wedge \bar{u}~,\cr 
d\f~=~& -\fracm{1}{2}(v+\bar{v})~.\cr}\eqno(A.17)$$

\section*{AppendixB: Basic facts about theta-functions}

The first theta-function $\vq_1(z|\t)$  is defined by the series \cite{theta}
$$\eqalign{
 \vq_1(z) \equiv  \vq_1(z|\t) & = 
-i\sum^{+\infty}_{n=-\infty}(-1)^n\exp i \left\{
\left( n +\fracm{1}{2}\right)^2\p\t + (2n+1)z\right\} \cr
 & = 2\sum^{+\infty}_{n=0}(-1)^n q^{(n+1/2)^2}\sin(2n+1)z~,\quad
q=e^{i\p\t}~, \cr} \eqno(B.1)$$
where $\t$ is regarded as the fundamental complex parameter, whose imaginary
part must be positive, $q$ is called the nome of the theta-function,
$\abs{q}<1$, and $z$ is the complex variable. The other theta-functions are
defined by \cite{theta}
$$\eqalign{
\vq_2(z|\t) & = ~\vq_1(z+\fracm{1}{2}\p)|\t) =
 \sum^{+\infty}_{n=-\infty} q^{(n+1/2)^2}e^{i(2n+1)z} \cr
& =  2\sum^{+\infty}_{n=0} q^{(n+1/2)^2}\cos(2n+1)z~,\cr}\eqno(B.2)$$
$$\eqalign{
\vq_3(z|\t) & = ~\vq_4(z+\fracm{1}{2}\p)|\t) =
 \sum^{+\infty}_{n=-\infty} q^{n^2}e^{2inz} \cr
& = 1+ 2\sum^{+\infty}_{n=1} q^{n^2}\cos 2nz~,\cr} \eqno(B.3)$$
and
$$ \vq_4(z|\t) = \sum^{+\infty}_{n=-\infty} (-1)^n q^{n^2}e^{2inz} =
1+ 2\sum^{+\infty}_{n=1} (-1)^n q^{n^2}\cos 2nz~.\eqno(B.4)$$

The identities \cite{theta}
$$ \vq^4_3(0) = \vq^4_2(0)+\vq^4_4(0)~,\eqno(B.5)$$
$$ \vq_1'(0) = \vq_2(0)\vq_3(0)\vq_4(0)~,\eqno(B.6)$$
and
$$ \fracmm{\vq_1'''(0)}{\vq_1'(0)}= 
\fracmm{\vq_2''(0)}{\vq_2(0)}+\fracmm{\vq_3''(0)}{\vq_3(0)}+
\fracmm{\vq_4''(0)}{\vq_4(0)}~~~,\eqno(B.7)$$
where the primes denote differentiation with respect to $z$, may be used to
 rewrite eq.~(6.4) to other equivalent forms (cf.~\cite{hit,dub,bk}).


\begin{thebibliography}{99}

\bibitem{strings} M. B. Green, J. H.  Schwarz and E. Witten, 
{\it Superstring Theory}, in two Volumes, Cambridge University Press, 1987;\\
J. Polchinski, {\it String Theory}, in two Volumes, Cambridge University 
Press, 1998;\\
S. V. Ketov, {\it Introduction to Quantum Theory of Strings 
and Superstrings}, Nauka Publishers, 1990 (in Russian) 
\bibitem{bbs} K. Becker, M. Becker and A. Strominger,  \np{456}{95}{130}
\bibitem{bb} K. Becker and M. Becker, \np{551}{99}{102} 
\bibitem{ka} S. Kachru and J. McGreevy, Phys. Rev. {\bf D61} (2000) 026001
\bibitem{bw} J. Bagger and E. Witten, \np{222}{83}{1}
\bibitem{w1} E. Witten, \np{474}{96}{343}
\bibitem{one} A. Strominger, \pl{421}{98}{139}
\bibitem{wp} B. de Wit and A. van Proeyen, \np{245}{84}{89};\\
B. de Wit, P. G. Lauwers  and A. van Proeyen, \np{255}{85}{569}
\bibitem{fcg} S. Cecotti, S. Ferrara and L. Girardello, \ijmp{4}{89}{2475}
\bibitem{aspin} P. S. Aspinwall, {\bf JHEP} 9804 (1998) 019  
\bibitem{h2d} S. Kachru and C. Vafa, \np{450}{95}{69}; \\
S. Ferrara, J. A. Harvey, A. Strominger and C. Vafa, \pl{361}{95}{59}
\bibitem{hm} J. A. Harvey and G. Moore, {\it Superpotentials and membrane
 instantons}, hep-th/9907026
\bibitem{gs} M. Gutperle and M. Spalinski, JHEP {\bf 0006} (2000) 037
\bibitem{gs2} M. Gutperle and M. Spalinski, {\it Supergravity instantons for 
N=2 hypermultiplets}, hep-th70010192
\bibitem{din} S. V. Ketov, {\it Gravitational dressing of D-insantons},
hep-th/0010255; to appear in Phys. Lett. B.
\bibitem{fsh} S. Ferrara and S. Sabharwal, \cqg{6}{89}{L77}
\bibitem{ger} H. G\"unther, C. Herrmann and J. Louis, Fortschr. Phys.
{\bf 48} (2000) 119
\bibitem{ova} H. Ooguri and C. Vafa, \prl{77}{96}{3298}
\bibitem{wit2} E. Witten, \pl{155}{85}{151}
\bibitem{wit} E. Witten, \np{500}{97}{3} 
\bibitem{ecole} E. J. Cremmer, B. Julia and J. Scherk, \pl{76}{78}{409} 
\bibitem{town} P. K. Townsend, {\it M-theory from its superalgebra}, in the
Proceedings of the NATO Advanced Study Institute `Strings, Branes and 
Dualities', Cargese, 1997, p.~141; hep-th/9712004
\bibitem{bst} E. Bergshoeff, E. Sezgin and P. K. Townsend, \pl{189}{87}{75}
\bibitem{guv} R. G\"uven, \pl{276}{92}{49}
\bibitem{pag} D. N. Page, \pr{28}{83}{2976}
\bibitem{tens} J. H. Schwarz, \pl{367}{96}{97}
\bibitem{tens2} S. P. de Alwis, \pl{388}{96}{291}
\bibitem{r4} M. B. Green and M. Gutperle, \np{498}{97}{195};\\
I. Antoniadis, S. Ferrara, R. Minasian and K. S. Narain, 
\np{507}{97}{571}
\bibitem{cand} P. Candelas, X. C. de la Ossa, P. S. Green and L. Parkes,
\np{359}{91}{21}
\bibitem{dew} B. de Wit, and A. van Proeyen, \pl{252}{90}{221};\\
B. de Wit, F. Vanderseypen and A. van Proeyen, \np{400}{93}{463}
\bibitem{lebrun} C. R. LeBrun, J. Diff. Geom. {\bf 34} (1991) 223
\bibitem{besse} A. L. Besse, {\it Einstein Manifolds}, Springer-Verlag, 1987
\bibitem{ati} M.F. Atiyah and N.J. Hitchin: {\it The Geometry and Dynamics 
of Magnetic Monopoles}, Princeton University Press, 1988   
\bibitem{bf} C. P. Boyer and J. D. Finley, J. Math. Phys. {\bf 23} (1982) 1126
\bibitem{sh} N. Seiberg and S. Shenker, \pl{388}{96}{521}
\bibitem{ward} M. Saveliev,  \cmp{121}{89}{283};\\
R. S. Ward, \cqg{7}{90}{L95}
\bibitem{lez} A. N. Leznov, Theor. Math. Phys. {\bf 117} (1998) 1194
\bibitem{nlsm} S. V. Ketov, {\it Quantum Non-Linear Sigma-Models},
Springer-Verlag, 2000, Ch.~4
\bibitem{fef} C. L. Fefferman, Ann. Math. {\bf 103} (1976) 395
\bibitem{bab} R. Britto-Pacumio, A. Strominger and A. Volovich, JHEP
 {\bf 9911} (1999) 013
\bibitem{tod} K. P. Tod, Twistor Newsletter {\bf 39} (1995) 19
\bibitem{cftbook} S. V. Ketov, {\it Conformal Field Theory}, World Scientific, 
1995, Sect.~IX.2
\bibitem{witk} B. de Wit, B. Kleijn and S. Vandoren, Nucl. Phys. 
{\bf B568} (2000) 475
\bibitem{bak} I. Bakas and K. Sfetsos, \ijmp{12}{97}{2585} 
\bibitem{bak2} I. Bakas, Fortschr. Phys. {\bf 48} (2000) 9 
\bibitem{ol} D. Olivier, Gen. Rel. Grav. {\bf 23} (1991) 1349
\bibitem{k2} S. V. Ketov, \pl{469}{99}{136}
\bibitem{sw} N. Seiberg and E. Witten, {\it Gauge dynamics and compactification
to three dimensions}, hep-th/9607163
\bibitem{tod3} K. P. Tod, Phys. Lett. {\bf 190A} (1994)
\bibitem{hit} N. J. Hitchin, J. Diff. Geom. {\bf 42} (1995) 30
\bibitem{halphen} G.-H. Halphen, Sur un syst\`eme d'\'equations 
diff\'erentielles, C. R. Acad. Sci. Paris {\bf 92} (1881) 1101
\bibitem{abc} M. J. Ablowitz and P. A. Clarkson, {\it Solitons, Non-Linear 
Evolution Equations and Inverse Scattering}, Cambridge University Press, 1991
\bibitem{int} L. J. Mason and N. M. J. Woodhouse, {\it Integrability, 
Self-Duality, and Twistor Theory}, Clarendon Press, 1996
\bibitem{pp} H. Pedersen and Y. S. Poon, \cqg{7}{90}{1707}
\bibitem{m3} R. Maszczyk, L. J. Mason and N. M. J. Woodhouse, \cqg{11}{94}{65}
\bibitem{theta} D. F. Lawden, {\it Elliptic Functions and Applications},
Springer-Verlag, 1980
\bibitem{dub} B. A. Dubrovin, Funct. Anal. Appl. {\bf 24} (1990) 280
\bibitem{bk} M. V. Babich and D. A. Korotkin, Lett. Math. Phys. {\bf 46} 
(1998) 323
\bibitem{cal} C. Callan, J. A. Harvey and A. Strominger, \np{367}{91}{60}
\bibitem{zam2} A. B. Zamolodchikov, JETP Lett. {\bf 43} (1986) 731
\bibitem{last} S. V. Ketov, {\it Exact renormalization flow and domain 
walls from holography}, hep-th/0009187; to appear in Nucl. Phys. B.
\bibitem{hit2} N. J. Hitchin, {\it A new family of Einstein metrics}, in the
Proceedings of the Pisa Conference in Honour of E. Calabi, Cambridge University
Press, 1995
\bibitem{dan} A. Dancer, {\it A family of gravitational instantons}, Cambridge
preprint DAMTP 92--13 (1992), unpublished 
\bibitem{tod2} K. P. Tod, \cqg{12}{95}{1535}
\bibitem{cp2} G. W. Gibbons and C. N. Pope, \cmp{61}{78}{239}.

\end{thebibliography}
\end{document}